# Pattern Analysis of Tandem Repeats in Nlrp1

Sim-Hui. Tee[1]

1. Multimedia University, Cyberjaya, 63100 Malaysia  e-mail: shtee@mmu.edu.my

**ABSTRACT**

Pattern analysis of tandem repeats in gene is an indispensable computational approach to the understanding of the gene expression and pathogenesis of diseases. This research applied a computational motif model and database techniques to study the distribution of tandem repeats in Nlrp1 gene, which is a critical gene to detect the invading pathogens in the immunologic mechanisms. The frequency of tandem repeats in Nlrp1 gene was studied for mono-, di-, tri-, and tetranucleotides. Mutations of Nlrp1 gene were analyzed to identify the insertion, deletion, and substitution of nucleotides. The results of this research provide a basis for future work in computational drug design and biomedical engineering in tackling diseases associated with immune system.

## 1. INTRODUCTION

Tandem repeats are a stretch of nucleotides that repeat in a consecutive manner. They are ubiquitous in genome and their mutated forms are responsible for diseases [1]. Understanding of the patterns of tandem repeat may provide insights into the pathogenesis and gene expression patterns. The advancement of computational tools such as computer algorithms, databases and web servers provide an efficient avenue to analyze enormous volume of genetic data, such as the phenomenon of tandem repeat in large genome.

This research aims at analyzing the patterns of tandem repeat in Nlrp1 gene. This gene is known to play a critical role in immune system of human [2]. Its protein product, which is NLRP1 inflammasome, can detect bacteria and toxins in a signaling pathway that leads to the release of pro-inflammatory cytokines [3]. NLRP1 consists of an N-terminal pyrin domain (PYD), a central NATCH domain, a leucine-rich repeat domain (LRR), a FIIND and a CARD domain [4]. The release of the pro-inflammatory cytokines which is mediated by inflammasome triggers adaptive immune response to eliminate the invading pathogens. The study of the patterns of tandem repeats may provide insights into therapeutic strategies related to immunologic diseases.

## 2. METHODS

The nucleotide sequence of Nlrp1 gene was obtained from the GenBank of the National Center for Biotechnology Information. UCSC genome browser [5] was used to locate the locus of Nlrp1 on the chromosome. Besides, KEGG database [6] was used to extract the signaling pathways for Nlrp1 encoded inflammasome. The statistically expected tandem repeats were calculated based on mreps algorithm [7]. The motif model [8] was constructed by distinguishing non-motif from tandem repeat motif, as shown below.

$$\Theta = \{\Theta^B, \Theta^M\} = \begin{bmatrix} \theta^B_{-,0} \theta^M_{-,1} \theta^M_{-,2} \ldots \theta^M_{-,w} \\ \theta^B_{a1,0} \theta^M_{a1,1} \theta^M_{a1,2} \ldots \theta^M_{a1,w} \\ \theta^B_{a2,0} \theta^M_{a2,1} \theta^M_{a2,2} \ldots \theta^M_{a2,w} \\ \ldots \quad \ldots \quad \ldots \\ \theta^B_{aj,0} \theta^M_{aj,1} \theta^M_{aj,2} \ldots \theta^M_{aj,w} \end{bmatrix} \quad (1)$$

where $\Theta^B$ is non-motif and $\Theta^M$ is motif. Let $S^{ij}$ be the subsequence of length $W$ at position $j$ in a sequence $i$. Let $a$ be the symbol that occurs at a position $k$ ($1 \leq k \leq W$) of either $\Theta^M$ or $\Theta^B$. Given that $L$ is a non-empty set for the length of nucleotide sequence, the conditional probabilities that $S^{ij}$ is found using the motif model are [8]:

$$P_M(S^{ij}) = \prod_{k=1}^{W} \prod_{a=1}^{L} (\theta_{ak}^M)^{I(S_{i,j+k-1}=a)} \quad (2)$$

We derive the conditional probabilities that $S^{ij}$ to be found using the non-motif model from [8]:

$$P_B(S^{ij}) = \prod_{k=1}^{W} \prod_{a=1}^{L} (\theta_{a0}^B)^{I(S_{i,j+k-1}=a)} \quad (3)$$

We calculated the motif occurrence probability $Z$ at position $j$ in sequence $i$ using formula (4), where $\lambda$ is the prior probability of motif occurrence [8].

$$Z_{ij} = \frac{\lambda P_M S^{ij}}{(\lambda P_M S^{ij}) + (1-\lambda) P_B S^{ij}} \quad (4)$$

$S^{ij}$ is taken to be a genuine motif hit when the equation (5) is fulfilled [8], as shown below.

$$\log(P_M(S^{ij})/P_B(S^{ij})) \geq \log[(1-\lambda)/\lambda] \quad (5)$$

Importantly, due to the complexity of motif, the pseudo-count of motif is likely to happen in a gene. We identified pseudo-count as below [8]:

$$psc_{ak} = \sum_{b=1}^{L} \theta_{b,k+1}^M P_{a/b} \quad (6)$$

where $P_{a/b}$ is the BLOSUM substitution probability for amino acid $a$ from the observed amino acid $b$. Pseudo-count of motif will be excluded from the final results. For the identified tandem repeats, relative frequency was used to analyze the total repeat per 100 bases in the nucleotide sequence of Nlrp1 gene. Besides, COSMIC database [21] was used to mine the mutation information for Nlrp1 gene.

## 3. RESULTS AND DISCUSSION

Nlrp1 gene has a length of 5623 base pairs (bp) with an even distribution of nucleotide composition (A=23.97%, T=21.16%, G=26.96%, C=27.90%). It is located on chromosome 17, as shown in Figure. 1 below.

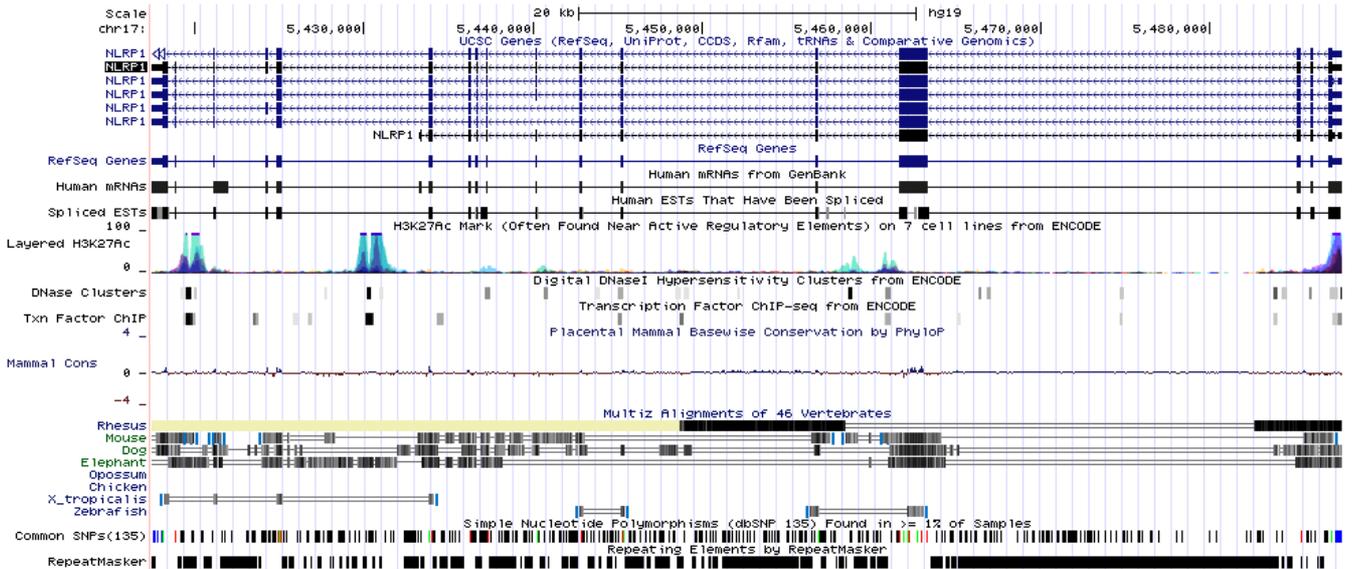

**Figure. 1: Nlrp1 gene on chromosome 17**

The bottom panel of Figure 1 demonstrates that there are extensive repeating elements in Nlrp1. We found that there are altogether 336 tandem repeats (in the form of mono-, di-, tri-, and tetranucleotide) on the gene. A total of 190 bp of interspersed repeats were found, amounting to 3.38% of the total nucleotides. The frequency of mono-, di-, tri-, and tetranucleotide tandem repeats is given in Table 1.

**Table 1: The frequency of tandem repeats in Nlrp1 gene**

| Tandem repeat | Occurrence frequency | Relative frequency |
|---|---|---|
| Mononucleotide | 90 | 1.60 |
| Dinucleotide | 187 | 3.33 |
| Trinucleotide | 51 | 0.91 |
| Tetranucleotide | 8 | 0.14 |

We also analyzed the distribution of tandem repeats on Nlrp1 gene, using a window size of 1000 bp. The distribution pattern is shown in Table 2. Where the sequence repeats were found spanning across the window (e.g., start locus at 999, end locus at 1004 for a GA repeat), such repeats are categorized according to the maximal nucleotide number. For example, in the case where a dinucleotide GA tandem repeats are spanning from loci 999 to 1004, they will be categorized under the window 1-2kbp. We observed a somehow even distribution of tandem repeats across the loci of Nlrp1 gene, as shown in Table 2. The number of tandem repeats at the last locus window (>5kbp) is lower than other regions primarily because the nucleotide length for this region is shorter (only 623 bp) than other gene region (1000 bp).

**Table 2: The distribution of tandem repeats (occurrence frequency)**

| Repeats | <1kbp | 1-2kbp | 2-3kbp | 3-4kbp | 4-5kbp | >5kbp |
|---|---|---|---|---|---|---|
| Mono- | 19 | 20 | 13 | 9 | 11 | 18 |
| Di- | 28 | 32 | 35 | 37 | 38 | 17 |
| Tri- | 13 | 12 | 13 | 6 | 3 | 4 |
| Tetra- | 2 | 1 | 1 | 1 | 2 | 1 |
| Total | 62 | 65 | 62 | 53 | 54 | 40 |

Though dinucleotide tandem repeats are the most abundant repeat type in Nlrp1 gene, the CG tandem repeat was not found. The CG/GC repeat has a frequency of 5 occurrences (2.67%) in the total of 187 dinucleotide tandem repeats, which is quite low as compared to the percentage of CG/GC (64.74%) in the total dinucleotide tandem repeats in Herpes simplex virus type 1 (HSV-1) genome [9]. However, the percentage of AG/GA occurrence in our study is 33.16%, whereas it was reported 4.78% in HSV-1 genome [9]. We also found a high percentage of CT/TC (28.88%) in Nlrp1 gene, while it was reported a low percentage (3.78%) in HSV-1 genome [9]. These variations in the abundance level of dinucleotide tandem repeats demonstrate the diverging genetic characteristics between human and viruses.

Besides, the abundance level and distribution of mononucleotide, trinucleotide and tetranucleotide tandem repeats in Nlrp1 gene are dissimilar to the findings observed in HSV-1 genome [9]. Because Ouyang et al. [9] used a measurement of per 1000 bp for the relative frequency (due to the larger genome size), and we used a measurement of per 100 bp (for small gene size), we converted Ouyang et al.'s relative frequency to our measurement unit for the purpose of comparison. In Ouyang et al's findings [9], the relative frequency is 0.436, 0.121, and 0.013 for mono-, tri-, and tetranucleotide tandem repeats, respectively. Ouyang et al's results showed that the occurrence of these tandem repeats is much lower in the genome of HSV-1 than in human Nlrp1 gene. Higher relative frequency of tandem repeat implies that the gene is more predisposed to the risk of mutation in the region of tandem repeat, therefore exposed to the phenotypic changes such as diseases.

We observed 25 tandem repeats which occur in the order higher than 4, as listed in Table 3.

**Table 3: The tandem repeats occur in the order >4**

| Repeat motif | Start locus | End locus |
|---|---|---|
| $(C)_5$ | 384 | 388 |
| $(C)_5$ | 524 | 528 |
| $(C)_5$ | 1082 | 1086 |
| $(C)_5$ | 1249 | 1253 |
| $(C)_5$ | 1276 | 1280 |
| $(C)_5$ | 4193 | 4197 |
| $(C)_5$ | 4390 | 4394 |
| $(G)_5$ | 1554 | 1558 |
| $(G)_5$ | 1607 | 1611 |
| $(G)_5$ | 2598 | 2602 |
| $(G)_5$ | 4075 | 4079 |
| $(G)_5$ | 5151 | 5155 |
| $(T)_5$ | 5504 | 5508 |
| $(T)_6$ | 5535 | 5540 |
| $(A)_5$ | 1363 | 1367 |
| $(A)_5$ | 1382 | 1386 |
| $(A)_5$ | 2168 | 2172 |
| $(A)_7$ | 2297 | 2303 |
| $(A)_5$ | 4948 | 4952 |
| $(A)_5$ | 5530 | 5534 |
| $(A)_7$ | 5558 | 5564 |
| $(A)_6$ | 5568 | 5573 |
| $(A)_6$ | 5575 | 5580 |
| $(A)_{11}$ | 5613 | 5623 |
| $(AG)_6$ | 1219 | 1230 |

From Table 3, all but one tandem repeats occur in the order higher than 4 are mononucleotide. The distribution pattern of high order tandem repeats is given in Table 4.

**Table 4: The distribution of high order (>4) tandem repeats**

| Order | <1kbp | 1-2kbp | 2-3kbp | 3-4kbp | 4-5kbp | >5kbp |
|---|---|---|---|---|---|---|
| 5 | 2 | 7 | 2 | 0 | 4 | 3 |
| 6 | 0 | 1 | 0 | 0 | 0 | 3 |
| 7 | 0 | 0 | 1 | 0 | 0 | 1 |
| >7 | 0 | 0 | 0 | 0 | 0 | 1 |
| Total | 2 | 8 | 3 | 0 | 4 | 8 |

We observed that most of the high order tandem repeats occur at loci region 1-2kbp and region >5kbp. These two regions are at the vicinity of 5' and 3' terminal region, mutation of which may have impacts on the transcriptional specificity (i.e., the binding affinity of transcription factors, RNA polymerases and enhancers). The observed mutations may either result in overexpression or down-regulation of a gene product [10-14]. The aberration in gene transcription and mutation in the exons may lead to diseases such as cancer and other intractable disorders [15-19]. However, the order of tandem repeats observed in Nlrp1 is far lesser than the repeat length observed in larger genetic region (e.g., chromosomes), where the tandem repeats can occur between different genes. Astolfi et al. [20] have reported the occurrence of tandem repeats in the order of greater than 50, such as $(ACC)_{82}$ and $(ATC)_{175}$ in human chromosome 22.

We used COSMIC database [21] to mine the mutation information for Nlrp1 gene. There were no deletion and insertion mutation found for Nlrp1 gene, yet the substitution mutations are quite extensive. Figure 2 illustrates the distribution of point substitution mutation for coding strand in Nlrp1 gene.

| Color | Mutation type | Mutant samples | Percentage |
|---|---|---|---|
| | A>C | 1 | 2.86% |
| | A>G | 1 | 2.86% |
| | A>T | 2 | 5.71% |
| | C>A | 2 | 5.71% |
| | C>T | 8 | 22.86% |
| | C>G | 1 | 2.86% |
| | G>A | 12 | 34.29% |
| | G>C | 0 | 0.00 |
| | G>T | 4 | 11.43% |
| | T>A | 1 | 2.86% |
| | T>C | 1 | 2.86% |
| | T>G | 2 | 5.71% |
| | Total | 32 | 100% |

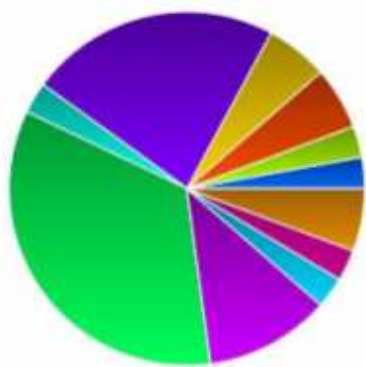

**Figure. 2: point substitution mutation in Nlrp1**

Figure 2 shows that a total of 32 mutant samples were found in Nlrp1 gene, among which the highest rate of mutation type being G>A nucleotide substitution (34.29%). C>T nucleotide substitution (22.86%) ranked the second highest rate of mutation type, followed with G>T substitution (11.43%). In addition, our analysis on the point substitution mutation for both coding and template strands demonstrated that C:G>T:A is the most common nucleotide substitution (57.14%), whereas C:G>G:C is the least common (2.86%). Mutation may result in aberration or loss-of-function in the gene [22-25], consequently results in the change of phenotype. Mutation in gene may culminate in various intractable diseases such as cancers [26-29], neurological diseases [30-31], coronary disease [32], infectious diseases [33-34], and developmental disorders [35]. Functional analysis on the mutation and its association to the tandem repeats should be carried out in future to understand the impact of polymorphism of tandem repeat on the gene expression.

## 4. CONCLUSION

Nlrp1 gene is known to play an essential role in immune system of human, aberration of which may lead to the dysregulation of pathogen sensing and diseases. This research has adopted a computational approach to investigate the patterns of tandem repeats in Nlrp1 gene. We observed a somehow even distribution of tandem repeats across the loci of Nlrp1 gene. Despite dinucleotide tandem repeats are the most abundant repeat type in Nlrp1 gene, the CG tandem repeat was not found. We used COSMIC database to mine the mutation information for Nlrp1 gene. The retrieved data show that there were no deletion and insertion mutation, while the substitution mutations are quite extensive in Nlrp1 gene. The results of this study provide insights for the computational drug design in pharmaceutical industry.

REFERENCES

1. R. Gemayel, M. D. Vinces, M. Legendre, and K.J. Verstrepen, "Variable tandem repeats accelerate evolution of coding and regulatory sequences," *Annual Review of Genetics*, vol. 44, pp. 445-477, 2010.
2. K. Schroder and J. Tschopp, "The Inflammasomes," *Cell*, vol. 140, pp. 821-832, 2010.
3. F.L. van de Veerdonk, M.G. Netea, C.A. Dinarello, and L. Joosten, "Inflammasome activation and IL-1β and IL-18 processing during infection," *Trends in Immunology*, vol. 32, pp. 110-116. 2011.
4. F. Martinon, A. Mayor, and J. Tschopp, "The inflammasomes: Guardians of the body," *Annual Reviews of Immunology*, Vol. 27, pp. 229-265, 2009.
5. P.A. Fujita, B. Rhead, A.S. Zweig, A.S. Hinrichs, D. Karolchik, M.S. Cline, M. Goldman, G.P. Barber, H. Clawson, A. Coelho, M. Diekhans, T.R. Dreszer, B.M. Giardine, R.A. Harte, J. Hillman-Jackson, F. Hsu, V. Kirkup, R.M. Kuhn, K. Learned, C.H. Li, L.R. Meyer, A. Pohl, B.J. Raney, K.R. Rosenbloom, K.E. Smith, D. Haussler, and W.J. Kent, "The UCSC Genome Browser database: update 2011", *Nucleic Acids Research*, vol. 39, pp. D876-D882, 2011.
6. M. Kanehisa, S. Goto, S. Kawashima, and A. Nakaya, "The KEGG databases at GenomeNet," *Nucleic Acids Research*, vol. 30, pp. 42-46, 2002.
7. R. Kolpakov, G. Bana, and G. Kucherov, "mreps: efficient and flexible detection of tandem repeats in DNA," *Nucleic Acids Research*, vol. 31, no. 13, pp. 3672-3678, 2003.
8. T. Le, T. Altman, and K. Gardiner, "HIGEDA: a hierarchical gene-set genetics based algorithm for finding subtle motifs in biological sequences", *Bioinformatics*, vol. 26, no. 3, pp. 302-309, 2010.
9. Q. Ouyang, X. Zhao, H. Feng, Y. Tian, D. Li, M. Li, and Z. Tan, "High GC content of simple sequence repeats in Herpes simplex virus type 1 genome," *Gene*, vol. 499, pp. 37-40, 2012.
10. P. Peidis, N. Voukkalis, E. Aggelidou, E. Georgatsou, M. Hadzopoulou-Cladaras, R. Scott, E. Nikolakaki, T. Giannakouros, "SAFB1 interacts with and suppresses the transcriptional activity of p53," *FEBS Letters*, vol. 585, pp. 78-84, 2011.
11. C. Le Goff, C. Mahaut, A. Abhyankar, W. Le Goff, V. Serre, A. Afenjar, A. Destrée, M. di Rocco, D. Héron, S. Jacquemont, S. Marlin, M. Simon, J. Tolmie, A. Verloes, J-L. Casanova, A. Munnich, and V. Cormier-Daire, "Mutations at a single codon in Mad homology 2 domain of SMAD4 cause Myhre syndrome," *Nature Genetics*, vol. 44, pp. 85-89, 2012.
12. A. Glaviano, C. Mothersill, C.P. Case, M.A. Rubio, R. Newson, and F. Lyng, "Effects of hTERT on genomic instability caused by either metal or radiation or combined exposure," *Mutagenesis*, vol. 24, pp. 25-33, 2009.
13. J. Chen, Z. Ma, X. Jiao, R. Fariss, W.L. Kantorow, M. Kantorow, E. Pras, M. Frydman, E. Pras, S. Riazuddin, S. Riazuddin, and J.F. Hejtmancik, "Mutations in FYCO1 cause autosomal-recessive congenital cataracts," *The American Journal of Human Genetics*, vol. 88, pp. 827-838, 2011.
14. M. Chotalia, S.A. Smallwood, N. Ruf, C. Dawson, D. Lucifero, M. Frontera, K. James, W. Dean, and G. Kelsey, "Transcription is required for establishment of germline methylation marks at imprinted genes," *Genes & Development*, vol. 23, pp. 105-117, 2009.
15. J. Vogt, B.J. Harrison, H. Spearman, J. Cossins, S. Vermeer, L. Cate, N.V. Morgan, D. Beeson, and E.R. Maher, "Mutation analysis of CHRNA1, CHRNB1, CHRND, and RAPSN genes in multiple pterygium syndrome/fetal akinesia patients," *The American Journal of Human Genetics*, vol. 82, pp. 222-227, 2008.
16. S. Sengupta, A. Shimamoto, M. Koshiji, R. Pedeux, M. Rusin, E.A. Spillare, J.C. Shen, L.E. Huang, N.M. Lindor, Y. Furuichi, and C.C. Harris, "Tumor suppressor p53 represses transcription of RECQ4 helicase," *Oncogene*, vol. 24, pp. 1738-1748, 2005.
17. H. Nilsen, Q. An, and T. Lindahl, "Mutation frequencies and AID activation state in B-cell lymphomas from Ung-deficient mice," *Oncogene*, vol. 24, pp. 3063-3066, 2005.
18. P.W. Ang, W.Q. Li, R. Soong, and B. Lacopetta, "BRAF mutation is associated with the CpG island methylator phenotype in colorectal cancer from young patients," *Cancer Letters*, vol. 273, pp. 221-224, 2009.
19. R. Dahse and H. Kosmehl, "Detection of drug-sensitizing EGFR exon 19 deletion mutations in salivary gland carcinoma," *British Journal of Cancer*, vol. 99, pp. 90-92, 2008.
20. P. Astolfi, D. Bellizzi, and V. Sgaramella, "Frequency and coverage of trinucleotide repeats in eukaryotes," *Gene*, vol. 317, pp. 117-125, 2003.
21. S.A. Forbes, N. Bindal, S. Bamford, C. Cole, C.Y. Kok, D. Beare, M. Jia, R. Shepherd, K. Leung, A. Menzies, J.W. Teague, P.J. Campbell, M.R. Stratton, and P.A. Futreal, "COSMIC: mining complete cancer genomes in the Catalogue of Somatic Mutations in Cancer," *Nucleic Acids Research*, vol. 39, pp. D945-D950, 2011.
22. G. Zifarelli and M. Pusch, "Conversion of the 2 $Cl^-/1\ H^+$ antiporter ClC-5 in a $NO_3^-/H^+$ antiporter by a single point mutation," *The EMBO Journal*, vol. 28, pp. 175-182, 2009.
23. M. Stražišar, V. Mlakar, and D. Glavač, "LATS2 tumour specific mutations and down-regulation of the gene in non-small cell carcinoma," *Lung Cancer*, vol. 64, pp. 257-262, 2009.
24. T. Soussi, "Advances in carcinogenesis: a historical perspective from observational studies to tumor genome sequencing and TP53 mutation spectrum analysis," *Biochimica et Biophysica Acta*, vol. 1816, pp. 199-208, 2011.
25. H.H. Nelson, B.C. Christensen, S.L. Plaza, J.K. Wiencke, C.J. Marsit, and K.T. Kelsey, "*KRAS* mutation, KRAS-LCS6 polymorphism, and non-small cell lung cancer," *Lung Cancer*, vol. 69, pp. 51-53, 2010.
26. M. Moghanibashi, P. Mohamadynejad, M. Rasekhi, A. Ghaderi, M. Mohammadianpanah, "Polymorphism of estrogen response element in TFF1 gene promoter is associated with an increased susceptibility to gastric cancer," *Gene*, vol. 492, pp. 100-103, 2012.
27. D. Dong, X. Gao, Z. Zhu, Q. Yu, S. Bian, and Y. Gao, "A 40-bp insertion/deletion polymorphism in the constitutive promoter of MDM2 confers risk for hepatocellular carcinoma in a Chinese population," *Gene*, vol. 497, pp. 66-70, 2012.
28. M. Gomes, A. Coelho, A. Araújo, A.L. Teixeira, R. Catarino, and R. Medeiros, "Influence of functional genetic polymorphism (-590C/T) in non-small cell lung cancer (NSCLC) development: The paradoxal role of IL-4," *Gene*, vol. 504, pp. 111-115, 2012.
29. F. Ye, J. Zhang, Q. Cheng, J. Shen, and H. Chen, "p53 codon 72 polymorphism is associated with occurrence of cervical carcinoma in the Chinese population," *Cancer Letters*, vol. 287, pp. 117-121, 2010.
30. E. Kesimci, A.B. Engin, O. Kanbak, and B. Karahalil, "Association between ABCB1 gene polymorphisms and fentanyl's adverse effects in Turkish patients undergoing spinal anesthesia," *Gene*, vol. 493, pp. 273-277, 2012.
31. J. Daborg, M. von Otter, A. Sjölander, S. Nilsson, L. Minthon, D. Gustafson, I. Skoog, K. Blennow, and H. Zetterberg, "Association of the RAGE G82S polymorphism with Alzheimer's disease," *Journal of Neural Transmission*, vol. 117, pp. 861-867, 2010.
32. M. Li, J. Shi, L. Fu, H. Wang, B. Zhou, and X. Wu, "Genetic polymorphism of MMP family and coronary disease susceptibility: A meta-analysis," *Gene*, vol. 495, pp. 36-41, 2012.
33. J. Guergnon and C. Combadière, "Role of chemokines polymorphisms in diseases," *Immunology Letters*, vol. 145, pp. 15-22, 2012.
34. L. Liang, Y-L. Zhao, J. Yue, J-F. Liu, M. Han, H. Wang, and H. Xiao, "Association of SP110 gene polymorphisms with susceptibility to tuberculosis in a Chinese population," *Infection, Genetics and Evolution*, vol. 11, pp. 934-939, 2011.
35. C. Le Goff, C. Mahaut, A. Abhyankar, W. Le Goff, V. Serre, A. Afenjar, A. Destrée, M. di Rocco, D. Héron, S. Jacquemont, S. Marlin, M. Simon, J. Tolmie, A. Verloes, J-L. Casanova, A. Munnich, and V. Cormier-Daire, "Mutations at a single codon in Mad homology 2 domain of SMAD4 cause Myhre syndrome," *Nature Genetics*, vol. 44, pp. 85-89, 2012.